\documentclass[preprint,12pt]{elsarticle}

\usepackage[utf8]{inputenc}
\usepackage[T1]{fontenc}
\usepackage{lmodern}
\usepackage{amssymb,amsthm,amsmath}
\usepackage{graphicx}
\usepackage{xcolor}
\usepackage{tikz}
\usepackage{booktabs}
\usepackage{multirow}
\usepackage{adjustbox}
\usepackage[ruled,vlined]{algorithm2e}
\usepackage{setspace}
\usepackage[hidelinks]{hyperref}
\SetKwInput{KwIn}{Input}
\SetKwInput{KwOut}{Output}
\newtheorem{thm}{Theorem}[section]

\newtheorem{defi}[thm]{Definition}

\newtheorem{example}[thm]{Example}
\numberwithin{equation}{section}
\journal{}

\begin{document}
\begin{spacing}{1.10}
\begin{frontmatter}
\title{Vertex--subgraph eigenvector centrality for connected induced subgraphs}
\author{Qingying Zhang}
\author{Haotian Zeng}
\author{Changjiang Bu}\ead{buchangjiang@hrbeu.edu.cn}
\address{School of Mathematical Sciences, Harbin Engineering University, Harbin, PR China}
\begin{abstract}
Eigenvector centrality evaluates the importance of vertices through adjacency relations,
while higher-order subgraph structures provide additional information for centrality analysis.
This motivates the simultaneous evaluation of vertices and subgraphs within a unified spectral framework.
In this paper, we propose vertex--subgraph eigenvector centrality (VSC), which jointly determines the centralities of vertices and
connected induced subgraphs of order three.
The vertex centrality depends on the centralities of the subgraphs containing it and those of the other vertices in these subgraphs, whereas the subgraph centrality depends on the centralities of its constituent vertices.
To distinguish induced paths $P_3$ from triangles $K_3$, the capacity of each subgraph is defined as its number of edges.
The resulting coupled nonlinear eigenvalue equations are represented by a fourth-order symmetric nonnegative tensor.
We prove that this tensor is weakly irreducible for every connected graph with at least three vertices, which guarantees the existence and uniqueness, up to scaling, of a positive Perron eigenvector.
Numerical experiments on four real-world networks compare VSC with classical centrality measures and show that capacity weighting has a greater effect on subgraph rankings than on vertex rankings.
\end{abstract}
\end{frontmatter}

\section{Introduction}
Centrality measures assign importance scores to vertices or subgraphs in a network on the basis of their structural roles. Common vertex-level measures include degree centrality \cite{albert2000error}, closeness centrality \cite{skibski2023closeness}, betweenness centrality \cite{barthelemy2004betweenness}, eigenvector centrality \cite{bonacich1972factoring}, and subgraph centrality \cite{estrada2005subgraph}. These measures have been used to investigate social and biological networks and other complex systems \cite{poulin2000dynamical,estrada2010generalized}.

Let $G$ be a connected graph, and let $A$ be the adjacency matrix of $G$.
Eigenvector centrality of $G$ is determined by the positive eigenvector $\mathbf{x}$ associated with the spectral radius $\rho(A)$ of $A$, satisfying
\begin{equation}\label{eq:classical-ec}
    \rho(A) x_i
    = \sum_{j=1}^{n} a_{ij}x_j,
    \qquad i\in[n].
\end{equation}
Equation~\eqref{eq:classical-ec} determines the score of a vertex from those of its adjacent vertices, with each incident edge contributing the score of its other endpoint.
Since an edge is a connected subgraph of order two, a natural extension is to incorporate connected subgraphs of higher order into the definition of vertex centrality.

One such extension is the $F$-subgraph eigenvector centrality introduced by Zhang et al.~\cite{zhang2026subgraph}.
For a prescribed connected graph $F$, this centrality is defined by the Perron eigenvector of a nonnegative tensor associated with the subgraphs of $G$ isomorphic to $F$.
The score of a vertex depends on the scores of the other vertices in these subgraphs, and the classical eigenvector centrality is recovered when $F=K_2$.
Subsequently, the $\alpha$-triangle eigenvector centrality~\cite{zhang2026alpha} incorporated both edge and triangle relations into a nonnegative tensor, with a parameter controlling their relative contributions.
Other approaches have defined vertex scores using triangle statistics~\cite{burkhardt2024triangle}, counts of prescribed families of subgraphs~\cite{bugedo2024family}, or tensor-based formulations of second-order adjacency relations~\cite{arrigo2020framework,xu2023two}.
In each of these approaches, subgraph structure contributes to the evaluation of vertices, but the subgraphs themselves are not assigned separate centrality scores.

Connected subgraphs may also be regarded as objects to be ranked. 
For example, Giscard and Wilson~\cite{giscard2018centrality} defined a centrality measure for cycles and more general subgraphs,
while Jiang et al.~\cite{jiang2023searching} ranked cycles using the variation of the Fiedler eigenvalue of the graph Laplacian. Song~\cite{song2024quantifying} also studied importance measures for specified higher-order substructures.
These methods assign scores to subgraphs, but do not define the vertex and subgraph scores simultaneously as components of a common positive eigenvector.

Coupled eigenvalue equations provide one way to define such joint scores.
The HITS algorithm~\cite{kleinberg1999authoritative} determines hub and authority vectors from two interdependent linear equations.
For multiplex networks, Tudisco et al.~\cite{tudisco2018node} defined node and layer eigenvector centralities.
Rahmede et al.~\cite{rahmede2018centralities} proposed a coupled system of equations to determine node and layer centralities simultaneously.
Bonacich et al.~\cite{bonacich2004hyperedges} studied multidimensional centrality in the presence of hyperedges.
Node and hyperedge centralities were subsequently defined through coupled nonlinear eigenvalue equations~\cite{tudisco2021node}.

Motivated by these considerations, we introduce vertex--subgraph eigenvector centrality (VSC) to determine the centralities of vertices and connected induced subgraphs of order three simultaneously.
The centrality of each vertex depends on the centralities of the subgraphs containing it and on the scores of the other vertices in those subgraphs.
Conversely, the centrality of each induced subgraph depends on the scores of its constituent vertices.
The number of edges in each subgraph is used as a capacity to distinguish $P_3$ from $K_3$.
These relations yield a system of coupled nonlinear eigenvalue equations, which we represent by a symmetric nonnegative tensor of order four.

The remainder of this paper is organized as follows.
Section~2 recalls some basic definitions and results from the spectral theory of nonnegative tensors.
Section~3 introduces the coupled vertex--subgraph centrality equations and constructs the corresponding fourth-order vertex--subgraph tensor.
The weak irreducibility of this tensor is established, and the vertex--subgraph eigenvector centrality (VSC) is defined in terms of its positive eigenvector.
Section~4 presents numerical experiments on an illustrative graph and four real-world networks to compare VSC with classical vertex centrality measures and examine the effects of capacity weighting on vertex and subgraph rankings.

\section{Preliminaries}
For a positive integer $n$, let $[n]=\{1,\ldots,n\}$.
Let $\mathbb R_+^n$ and $\mathbb R_{++}^n$ denote the nonnegative and positive orthants of $\mathbb R^n$, respectively.
For a vector $\mathbf{x}=(x_1,\ldots,x_n)^\mathrm{T}$ and a positive integer $r$, define $\mathbf{x}^{[r]}=(x_1^r,\ldots,x_n^r)^\mathrm{T}$.

Let $\mathcal A=(a_{i_1\cdots i_k})$ be a $k$th-order $n$-dimensional tensor, where $i_1,\ldots,i_k\in[n]$.
For $\mathbf{x}\in\mathbb R^n$, the vector $\mathcal A\mathbf{x}^{k-1}$ is defined by
\begin{equation}\label{eq:tensor-vector-product}
(\mathcal A\mathbf{x}^{k-1})_i
=
\sum_{i_2,\ldots,i_k=1}^n
a_{i i_2\cdots i_k}
x_{i_2}\cdots x_{i_k},
\quad i\in[n].
\end{equation}
The tensor $\mathcal A$ is said to be nonnegative if all its entries are nonnegative.

\begin{defi}\cite{qi2005eigenvalues,lim2005singular}
A scalar $\lambda\in\mathbb C$ is called an
eigenvalue of a $k$th-order $n$-dimensional
tensor $\mathcal A$ if there exists a nonzero
vector $\mathbf{x}\in\mathbb C^n$ such that
\begin{equation}\label{eq:tensor-eigenpair}
\mathcal A\mathbf{x}^{k-1}
=
\lambda\mathbf{x}^{[k-1]}.
\end{equation}
In this case, $\mathbf{x}$ is called an
eigenvector of $\mathcal A$ corresponding to
$\lambda$. The spectral radius of $\mathcal A$,
denoted by $\rho(\mathcal A)$, is defined as
the maximum modulus of its eigenvalues.
\end{defi}

For a nonnegative tensor $\mathcal A$, the
associated directed graph $D(\mathcal A)$
has vertex set $[n]$, with an arc $i\to j$
if there exist indices $i_2,\ldots,i_k\in[n]$
such that $a_{i i_2\cdots i_k}>0$ and
$j\in\{i_2,\ldots,i_k\}$
\cite{qi2017tensor}.
The tensor $\mathcal A$ is said to be
\emph{weakly irreducible} if $D(\mathcal A)$
is strongly connected
\cite{friedland2013perron}.
Otherwise, $\mathcal A$ is said to be
\emph{weakly reducible}.
Equivalently, $\mathcal A$ is weakly reducible
if there exists a nonempty proper subset
$I\subset[n]$ such that
$a_{i i_2\cdots i_k}=0$
whenever $i\in I$ and at least one of
$i_2,\ldots,i_k$ belongs to $[n]\setminus I$.

\begin{thm} \cite{friedland2013perron} \label{thm:tensor-pf}
If $\mathcal A$ is a nonnegative weakly irreducible tensor, then $\rho(\mathcal A)$ is a positive eigenvalue admitting a positive eigenvector. This vector is unique up to positive scaling.
\end{thm}

\section{Vertex--subgraph eigenvector centrality}\label{sec:model}
Let $G=(V,E)$ be a connected undirected graph with $V=[n]$ and $n\geq 3$.
Let $G[V_1],\ldots,G[V_m]$ denote all connected induced subgraphs of $G$ on three vertices,
where
\[
V_\alpha=\{i_\alpha,j_\alpha,k_\alpha\},
\qquad \alpha\in[m].
\]
Each $G[V_\alpha]$ is isomorphic to either the path $P_3$ or the triangle $K_3$.
Its \emph{capacity} is its number of edges:
\begin{equation}\label{eq:capacity}
 \kappa_\alpha=|E(G[V_\alpha])|=
 \begin{cases}2,&G[V_\alpha]\cong P_3,\\3,&G[V_\alpha]\cong K_3.\end{cases}
\end{equation}
Let $x_i>0$ be the centrality score of vertex $i$, and let $y_\alpha>0$ be the centrality score of $G[V_\alpha]$. Write $\mathbf x=(x_1,\ldots,x_n)^\mathrm{T}$ and $\mathbf y=(y_1,\ldots,y_m)^\mathrm{T}$.

For $i\in V_\alpha$, the contribution of $G[V_\alpha]$ to vertex $i$ is taken to be
\[
 \kappa_\alpha y_\alpha\prod_{j\in V_\alpha\setminus\{i\}}x_j.
\]
Thus the contribution involves both the importance of the induced subgraph and the scores of the other two vertices. Conversely, the contribution determining the centrality of $G[V_\alpha]$ is
\[
 \kappa_\alpha\prod_{i\in V_\alpha}x_i.
\]
Requiring the two classes of scores to reinforce each other yields the coupled system
\begin{equation}\label{eq:vsc-coupled}
\begin{aligned}
 \lambda x_i^3
 &=\sum_{\substack{\alpha\in[m]\\i\in V_\alpha}}
 \kappa_\alpha y_\alpha
 \prod_{j\in V_\alpha\setminus\{i\}}x_j,
 &&i\in[n],\\[5pt]
 \lambda y_\alpha^3
 &=\kappa_\alpha\prod_{i\in V_\alpha}x_i,
 &&\alpha\in[m],
\end{aligned}
\end{equation}
where $\lambda>0$. Both equations are homogeneous of degree three. The vertex equation aggregates contributions from all three-vertex subgraphs containing $i$, whereas the subgraph equation determines $y_\alpha$ from the capacity and member-vertex scores. No position within an induced path is singled out: its three vertices enter the corresponding product symmetrically.

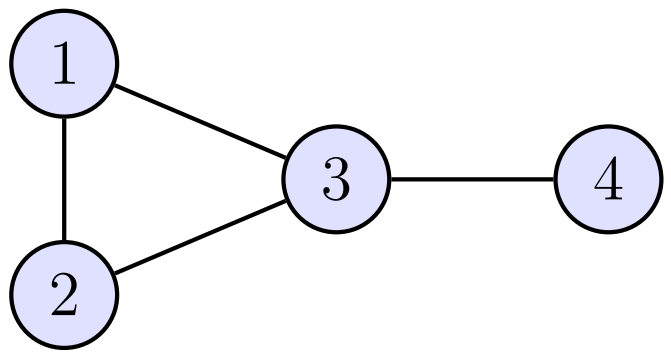
\begin{figure}[htbp]
\centering
\begin{tikzpicture}[scale=0.90, every node/.style={circle,draw=black,fill=blue!12,minimum size=7mm,inner sep=1pt},line width=0.8pt]
 \node (v1) at (0,1.55) {$1$};
 \node (v2) at (0,-0.15) {$2$};
 \node (v3) at (2,0.7) {$3$};
 \node (v4) at (4,0.7) {$4$};
 \draw (v1)--(v2)--(v3)--(v1);
 \draw (v3)--(v4);
\end{tikzpicture}
\caption{A graph with four vertices.}\label{fig:four-vertex-example}
\end{figure}

\begin{example}\label{ex:four-vertices}
Figure~\ref{fig:four-vertex-example} shows a triangle on $\{1,2,3\}$ with an additional edge $\{3,4\}$. The vertex sets of its three-vertex subgraphs are
\[
 V_1=\{1,2,3\},\quad V_2=\{1,3,4\},\quad V_3=\{2,3,4\},
\]
with capacities $\kappa_1=3$ and $\kappa_2=\kappa_3=2$. Since vertex $2$ belongs to $V_1$ and $V_3$, its equation in~\eqref{eq:vsc-coupled} is
\begin{equation*}
 \lambda x_2^3=3y_1x_1x_3+2y_3x_3x_4.
\end{equation*}
The first term is contributed by the triangle $G[V_1]$; the second comes from the induced path $G[V_3]$. For the triangle itself,
\begin{equation*}
 \lambda y_1^3=3x_1x_2x_3.
\end{equation*}
Thus, the score of vertex $2$ depends on the scores of the three-vertex subgraphs containing it, while the score of $G[V_1]$ depends on those of its member vertices. Together, these equations illustrate the mutual reinforcement underlying VSC.
\end{example}

Collect the vertex and subgraph scores in a single positive vector
\begin{equation}\label{eq:joint-vector}
 \mathbf z=(\mathbf x^\mathrm{T},\mathbf y^\mathrm{T})^\mathrm{T}\in\mathbb R_{++}^{n+m},
\end{equation}
so that $z_i=x_i$ for $i\in[n]$ and $z_{n+\alpha}=y_\alpha$ for $\alpha\in[m]$.
The first $n$ coordinates refer to vertices and the remaining $m$ coordinates refer to the induced subgraphs. The associated tensor is defined directly in terms of these indices.

\begin{defi}\label{def:vsc-tensor}
Let $G$ be a connected graph.
The fourth-order $(n+m)$-dimensional tensor
$\mathcal T_G=(t_{pqrs})$ is called the
\emph{vertex--subgraph tensor} of $G$,
where
\begin{equation}\label{eq:tensor-definition}
t_{pqrs}=
\begin{cases}
\dfrac{\kappa_\alpha}{6},
& \{p,q,r,s\}
=\{i_\alpha,j_\alpha,k_\alpha,n+\alpha\}
\text{ for some }\alpha\in[m],\\[5pt]
0, & \text{otherwise}.
\end{cases}
\end{equation}
\end{defi}

The vertex--subgraph tensor $\mathcal T_G$ is symmetric and nonnegative. For the graph in Fig.~\ref{fig:four-vertex-example}, $n=4$ and $V_3=\{2,3,4\}$; hence the index representing $G[V_3]$ is $n+3=7$. All $4!$ entries with the four distinct indices $(2,3,4,7)$ have value $\kappa_3/6=1/3$; in particular,
\[
 t_{2347}=t_{2374}=t_{7234}=\cdots=\frac{1}{3}.
\]
The factor $1/6$ in~\eqref{eq:tensor-definition} compensates for the $3!=6$ permutations of the remaining three indices when the first index is fixed. Consequently,
\begin{equation*}\label{eq:tensor-components}
 (\mathcal T_G\mathbf z^3)_i
 =\sum_{\substack{\alpha\in[m]\\i\in V_\alpha}}
 \kappa_\alpha y_\alpha\prod_{j\in V_\alpha\setminus\{i\}}x_j,
 i\in[n].
 \end{equation*}
 \begin{equation*}
 (\mathcal T_G\mathbf z^3)_{n+\alpha}
 =\kappa_\alpha\prod_{i\in V_\alpha}x_i,
 \alpha\in[m].
\end{equation*}
Thus~\eqref{eq:vsc-coupled} is exactly the tensor eigenvalue equation
\begin{equation}\label{eq:vsc-eigenvalue}
 \mathcal T_G\mathbf z^3=\lambda\mathbf z^{[3]}.
\end{equation}

\begin{thm}\label{thm:vsc-weak-irreducible}
Let $G$ be a connected graph on $n\geq3$ vertices. Then its vertex--subgraph tensor $\mathcal T_G$ is weakly irreducible.
\end{thm}
\begin{proof}
Consider the associated directed graph $D(\mathcal T_G)$ on $[n+m]$. For each $V_\alpha=\{i_\alpha,j_\alpha,k_\alpha\}$, the symmetry and positivity of the corresponding tensor entries imply that every pair of distinct indices in $\{i_\alpha,j_\alpha,k_\alpha,n+\alpha\}$ is joined by arcs in both directions. In particular, $D(\mathcal T_G)$ contains $i\to n+\alpha$ and $n+\alpha\to i$ for each $i\in V_\alpha$.

Every edge $uv\in E(G)$ is contained in some connected induced subgraph of order three. Indeed, since $G$ is connected and $n\geq3$, at least one endpoint of $uv$ has a neighbor $w\notin\{u,v\}$. Hence $G[\{u,v,w\}]$ is connected, so $\{u,v\}\subseteq V_\alpha$ for some $\alpha$. The preceding observation then gives the arcs $u\to v$ and $v\to u$ in $D(\mathcal T_G)$. As $G$ is connected, all indices in $[n]$ are mutually reachable by directed paths in $D(\mathcal T_G)$.

Finally, each subgraph index $n+\alpha$ is linked in both directions to every index in $V_\alpha$, and therefore belongs to the same strongly connected component as the vertex indices. It follows that $D(\mathcal T_G)$ is strongly connected. By the Theorem~\ref{thm:tensor-pf} in Section~2, $\mathcal T_G$ is weakly irreducible.
\end{proof}

By Theorems~\ref{thm:vsc-weak-irreducible} and~\ref{thm:tensor-pf}, the tensor $\mathcal T_G$ has a positive eigenvector corresponding to $\rho(\mathcal T_G)$, which is unique up to a positive scalar multiple.

\begin{defi}\label{def:vsc-centrality}
Let $G$ be a  connected graph on $n\geq3$ vertices. The vertex--subgraph eigenvector centrality (VSC) vector of $G$ is the unique positive vector $\mathbf z=(\mathbf x^\mathrm{T},\mathbf y^\mathrm{T})^\mathrm{T}\in\mathbb R_{++}^{n+m}$ satisfying
\begin{equation}\label{eq:vsc-definition}
 \mathcal T_G\mathbf z^3=\rho(\mathcal T_G)\mathbf z^{[3]},
 \qquad\|\mathbf z\|_4=1.
\end{equation}
The components of $\mathbf x$ and $\mathbf y$ are called the \emph{vertex VSC scores} and \emph{subgraph VSC scores}, respectively.
\end{defi}

\section{Numerical experiments}\label{sec:experiments}
We evaluate the vertex and subgraph VSC scores numerically. After describing the numerical procedure, we examine an eight-vertex graph and four real networks. The real-network experiments compare the vertex rankings produced by VSC with those of established measures and then assess the influence of subgraph capacity on vertex and subgraph rankings. Unless otherwise indicated, VSC scores are obtained from the normalized positive Perron eigenvector in Definition~\ref{def:vsc-centrality}.

\subsection{Numerical computation}
We first enumerate all connected induced subgraphs $G[V_1],\ldots,G[V_m]$ of $G$ on three vertices and compute their capacities $\kappa_\alpha$ according to~\eqref{eq:capacity}.
The tensor $\mathcal T_G$ is represented implicitly using these induced subgraphs, without explicitly constructing an $(n+m)^4$ array.
The positive eigenvector corresponding to $\rho(\mathcal T_G)$ is computed by the ZQW method~\cite{zhou2013efficient}.
We apply the shift $\mathcal B=\mathcal T_G+\mathcal I$, where $\mathcal I$ is the fourth-order
unit tensor.
The spectral radius of $\mathcal T_G$ is then estimated by subtracting one from the corresponding eigenvalue estimate of $\mathcal B$.
The resulting eigenvector is normalized to satisfy~\eqref{eq:vsc-definition}. The computational procedure is summarized in Algorithm~\ref{alg:VSC}.

\begin{algorithm}[htbp]
\footnotesize
\caption{Computation of VSC by the ZQW method}
\label{alg:VSC}

\KwIn{A connected graph $G$; tolerance $\varepsilon$;
maximum iteration number $T_{\max}$.}

\KwOut{VSC vector
$\mathbf z=(\mathbf x^\mathrm{T},
\mathbf y^\mathrm{T})^\mathrm{T}$
and spectral radius $\rho(\mathcal T_G)$.}

Enumerate all connected induced subgraphs
$G[V_1],\ldots,G[V_m]$ of $G$ on three
vertices and set
$\kappa_\alpha=|E(G[V_\alpha])|$
for $\alpha\in[m]$\;

Construct $\mathcal T_G$ implicitly and set
$\mathcal B=\mathcal T_G+\mathcal I$\;

Choose $\mathbf z^{(0)}>0$ with
$\|\mathbf z^{(0)}\|_1=1$\;

\For{$t=0,1,\ldots,T_{\max}-1$}{

  Compute
  $\mathbf w^{(t)}
  =\mathcal B(\mathbf z^{(t)})^3$\;

  $\displaystyle
  \underline{\lambda}^{(t)}
  =\min_{j\in[n+m]}
  \frac{w_j^{(t)}}{(z_j^{(t)})^3},
  \qquad
  \overline{\lambda}^{(t)}
  =\max_{j\in[n+m]}
  \frac{w_j^{(t)}}{(z_j^{(t)})^3}$\;

  \If{$\overline{\lambda}^{(t)}
  -\underline{\lambda}^{(t)}<\varepsilon$}{
    break\;
  }

  $\displaystyle
  \mathbf z^{(t+1)}
  =
  \frac{(\mathbf w^{(t)})^{[1/3]}}
       {\|(\mathbf w^{(t)})^{[1/3]}\|_1}$\;
}

$\displaystyle
{\rho}(\mathcal T_G)
=
\frac{\underline{\lambda}^{(t)}
+\overline{\lambda}^{(t)}}{2}-1$\;

Normalize $\mathbf z^{(t)}$ so that
$\|\mathbf z^{(t)}\|_4=1$\;

Set
$\mathbf x=(z_1^{(t)},\ldots,z_n^{(t)})^\mathrm T$
and
$\mathbf y=(z_{n+1}^{(t)},\ldots,
z_{n+m}^{(t)})^\mathrm T$\;

\Return{$\mathbf x,\mathbf y,
{\rho}(\mathcal T_G)$}\;

\end{algorithm}

\subsection{An  example}\label{subsec:eight-example}
\begin{figure}[htbp]
\centering
\includegraphics[width=0.65\linewidth]{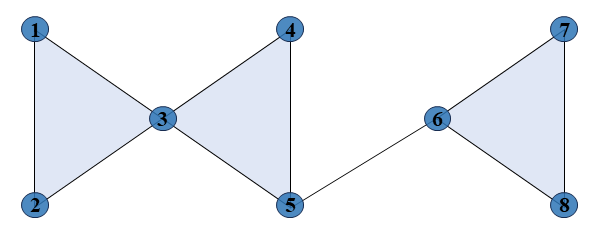}
\caption{A graph with $8$ vertices.}\label{fig:eight-vertex}
\end{figure}

Consider the graph in Fig.~\ref{fig:eight-vertex}. It contains both induced paths and triangles, allowing the two subgraph types to be compared within the same VSC model. The vertex rankings and subgraph rankings obtained from the same Perron eigenvector are given in Tables~\ref{tab:toy-vertices} and~\ref{tab:toy-subgraphs}, respectively.

\begin{table}[htbp]
\centering
\caption{Vertex ranking obtained by VSC for the graph in Fig.~\ref{fig:eight-vertex}.}\label{tab:toy-vertices}
\begin{tabular}{ccc}
  \toprule
  Rank & Vertex & VSC score\\
  \midrule
  1 & 3 & 0.6651\\
  2 & 5 & 0.6364\\
  3 & 4 & 0.5756\\
  4 & 6 & 0.5298\\
  5 & 1 & 0.5295\\
  5 & 2 & 0.5295\\
  7 & 7 & 0.3865\\
  7 & 8 & 0.3865\\
  \bottomrule
  \end{tabular}
\end{table}

\begin{table}[htbp]
\centering
\caption{Ranking of three-vertex subgraphs in
Fig.~\ref{fig:eight-vertex}.}
\label{tab:toy-subgraphs}
\begin{tabular}{ccccc}
\toprule
Rank & Subgraph & Type & Capacity & VSC score\\
\midrule
1  & $\{3,4,5\}$ & $K_3$ & 3 & 0.4653\\
2  & $\{1,2,3\}$ & $K_3$ & 3 & 0.4256\\
3  & $\{3,5,6\}$ & $P_3$ & 2 & 0.3954\\
4  & $\{1,3,5\}$ & $P_3$ & 2 & 0.3953\\
4  & $\{2,3,5\}$ & $P_3$ & 2 & 0.3953\\
6  & $\{1,3,4\}$ & $P_3$ & 2 & 0.3823\\
6  & $\{2,3,4\}$ & $P_3$ & 2 & 0.3823\\
8  & $\{4,5,6\}$ & $P_3$ & 2 & 0.3768\\
9  & $\{5,6,7\}$ & $P_3$ & 2 & 0.3299\\
9 & $\{5,6,8\}$ & $P_3$ & 2 & 0.3299\\
11  & $\{6,7,8\}$ & $K_3$ & 3 & 0.3199\\
\bottomrule
\end{tabular}
\end{table}

The three highest-ranked vertices are $3$, $5$, and $4$, with scores $0.6651$, $0.6364$, and $0.5756$, respectively. In particular, vertex $3$ is contained in both triangles $\{1,2,3\}$ and $\{3,4,5\}$ and in several induced paths. Its score is therefore reinforced by multiple three-vertex subgraphs and their other member vertices, as expressed by the vertex equation in~\eqref{eq:vsc-coupled}.

The two highest-ranked subgraphs are the triangles $\{3,4,5\}$ and $\{1,2,3\}$, with scores $0.4653$ and $0.4256$. However, the path induced by $\{3,5,6\}$ ranks third with score $0.3954$, whereas the triangle $\{6,7,8\}$ ranks last with score $0.3199$. The triangle $\{6,7,8\}$ includes vertices $7$ and $8$, which have the two lowest vertex scores. These comparisons demonstrate that subgraph ranking is not determined by isomorphism type alone.

Indeed, the second equation in~\eqref{eq:vsc-coupled} gives
\begin{equation}\label{eq:subgraph-score-comparison}
 \frac{y_\alpha}{y_\beta}
 =\left(\frac{\kappa_\alpha\prod_{i\in V_\alpha}x_i}
 {\kappa_\beta\prod_{j\in V_\beta}x_j}\right)^{1/3}.
\end{equation}
If $G[V_\alpha]\cong P_3$ and $G[V_\beta]\cong K_3$, the path outranks the triangle precisely when its product of member-vertex scores is greater than $3/2$ times the product for the triangle. Thus, capacity distinguishes the internal connection patterns, but the scores of their member vertices can reverse the order induced by capacity alone.

\subsection{Vertex rankings on real networks}\label{subsec:real-networks}
We use four real networks: the Karate club social network \cite{zachary1977information}, a C.~elegans network from the Network Data Repository \cite{rossi2015network}, the Watts power-grid network \cite{watts1998collective}, and the Facebook pages network (Fb-pages) \cite{rozemberczki2019gemsec}. 

We compare the vertex rankings obtained by VSC with those given by four classical centrality measures:
degree centrality (DC), eigenvector centrality (EC), triangle centrality (TC), and betweenness centrality (BC). Their definitions are briefly recalled below.

Degree centrality (DC) measures the number of neighbors of a vertex and is defined by
\[
DC(i)=d_i=|N(i)|,
\]
where $N(i)$ is the neighborhood of vertex $i$ \cite{albert2000error}.

Eigenvector centrality (EC) is determined by the positive eigenvector $\mathbf x$
corresponding to the spectral radius of the adjacency matrix $A$~\cite{bonacich1972factoring},
satisfying
\[
A\mathbf x
=\rho(A)\mathbf x.
\]
The centrality score of vertex $i$ is $EC(i)=x_i$.

Triangle centrality (TC)~\cite{burkhardt2024triangle} evaluates vertices using the numbers of triangles
containing them and their neighbors.
Let $\tau(i)$ denote the number of triangles containing vertex $i$, and let $\tau(G)$ be the total number of triangles in $G$.
Define
\[
N_\triangle(i)
=\{j\in N(i):\{i,j\}
\text{ belongs to a triangle}\},
\]
and $N_\triangle[i]=N_\triangle(i)\cup\{i\}$. Then, for $\tau(G)>0$,
\[
TC(i)
=
\frac{
\displaystyle
\frac{1}{3}\sum_{j\in N_\triangle[i]}\tau(j)
+
\sum_{j\in N(i)\setminus N_\triangle(i)}\tau(j)
}{\tau(G)}.
\]

Betweenness centrality (BC)~\cite{freeman1991centrality} measures the proportion of shortest paths
between pairs of vertices that pass through a given vertex. It is defined by
\[
BC(i)
=
\sum_{\substack{s<t\\s,t\neq i}}
\frac{\sigma_{st}(i)}{\sigma_{st}},
\]
where $\sigma_{st}$ is the number of shortest paths between vertices $s$ and $t$, and $\sigma_{st}(i)$ is the number of such paths
passing through vertex $i$ as an internal vertex.

Table~\ref{tab:network} reports network sizes and the numbers of three-vertex subgraphs. Here $m_{P_3}$ and $m_{K_3}$ denote the respective numbers of induced paths and triangles, so $m=m_{P_3}+m_{K_3}$. The average clustering coefficient $\langle C\rangle$ and the ratio $m_{K_3}/m$ describe different aspects of local closure. In particular, triangles form only $10.3\%$, $4.5\%$, $3.7\%$, and $5.7\%$ of all three-vertex subgraphs in Karate, C.~elegans, Watts, and Facebook, respectively. This class imbalance is relevant when interpreting the subgraph rankings.

\begin{table}[htbp]
\centering
\caption{Basic topological statistics of the four real networks.}
\label{tab:network}
\begin{adjustbox}{max width=\textwidth}
\begin{tabular}{lrrrrrrr}
\toprule
Network&$n$&$|E|$&$m$&$m_{P_3}$&$m_{K_3}$&$\langle C\rangle$&$m_{K_3}/m$\\
\midrule
Karate&34&78&438&393&45&0.571&0.103\\
C.~elegans&453&2025&72605&69321&3284&0.647&0.045\\
Watts&4941&6594&17631&16980&651&0.080&0.037\\
Fb-pages&14113&52126&984703&928698&56005&0.239&0.057\\
\bottomrule
\end{tabular}
\end{adjustbox}
\end{table}

\subsubsection{Spearman rank correlations}
\begin{figure}[htbp]
\centering
\includegraphics[width=0.86\linewidth]{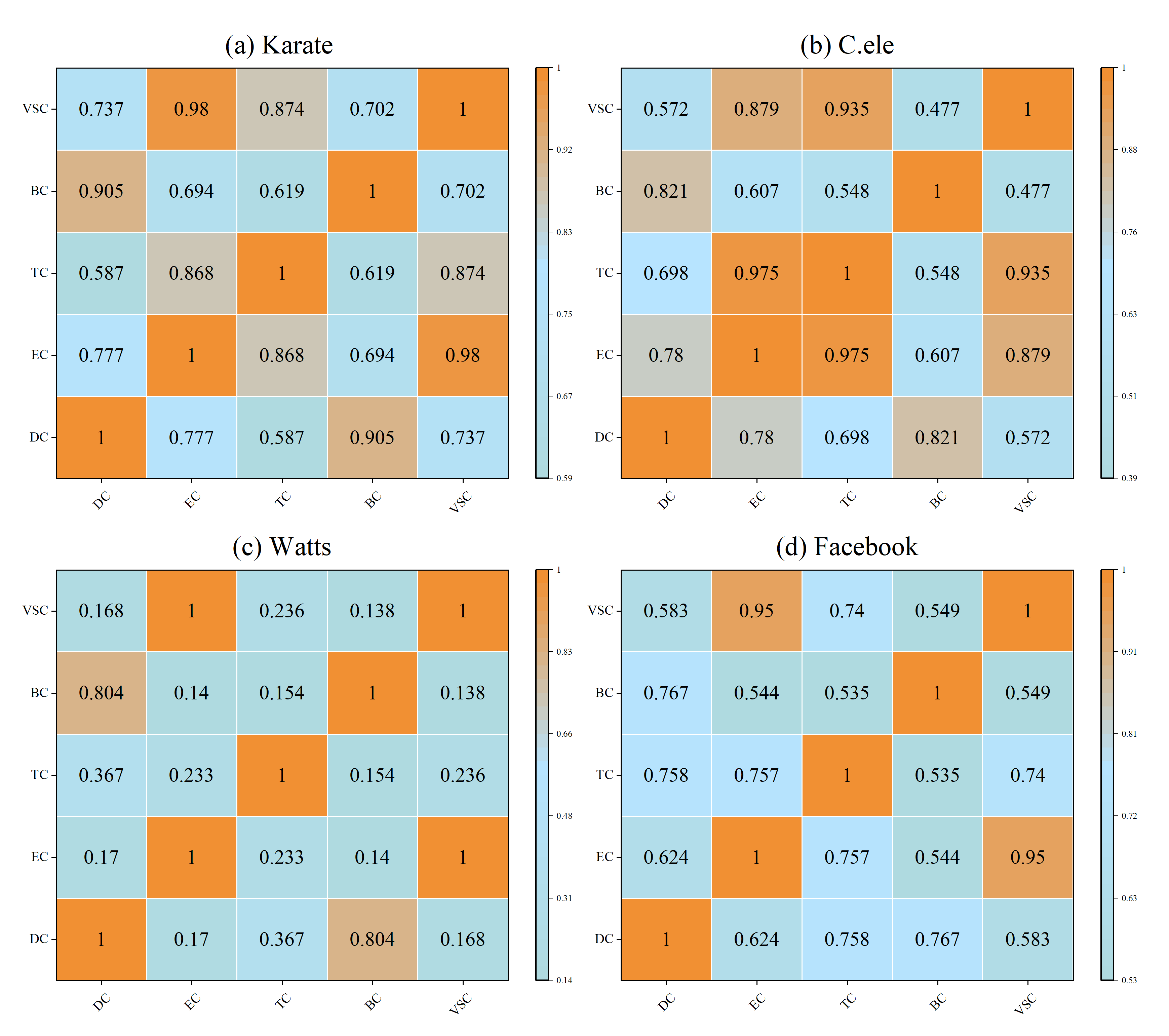}
\caption{Spearman rank correlations between VSC and four classical vertex-centrality measures on the four networks.}\label{fig:spearman}
\end{figure}

To compare whole-vertex rankings, we use Spearman's rank correlation coefficient \cite{spearman1987proof}. A coefficient close to $1$ indicates strong agreement in the relative ordering of vertices, but does not imply that the same vertices occupy the highest positions. Figure~\ref{fig:spearman} presents the correlations between VSC and the four comparison measures.

The VSC rankings are positively correlated with the classical rankings, although the strength of correlation depends on the network and the comparator. In Karate, the VSC--EC correlation is $0.980$, compared with $0.737$ for VSC--DC and $0.702$ for VSC--BC. For C.~elegans, VSC is more strongly correlated with TC ($0.935$) and EC ($0.879$) than with DC or BC. The VSC--EC correlation is reported as $1.000$ to three decimal places in Watts, whereas the corresponding correlations with DC, TC, and BC are $0.168$, $0.236$, and $0.138$. In Facebook, VSC--EC correlation reaches $0.950$. These contrasts indicate that agreement between VSC and a conventional vertex measure is network-dependent. They do not imply that one ranking is intrinsically more accurate than another.

\subsubsection{Top-\texorpdfstring{$k\%$}{k-percent} Jaccard overlap}
\begin{figure}[htbp]
\centering
\includegraphics[width=0.86\linewidth]{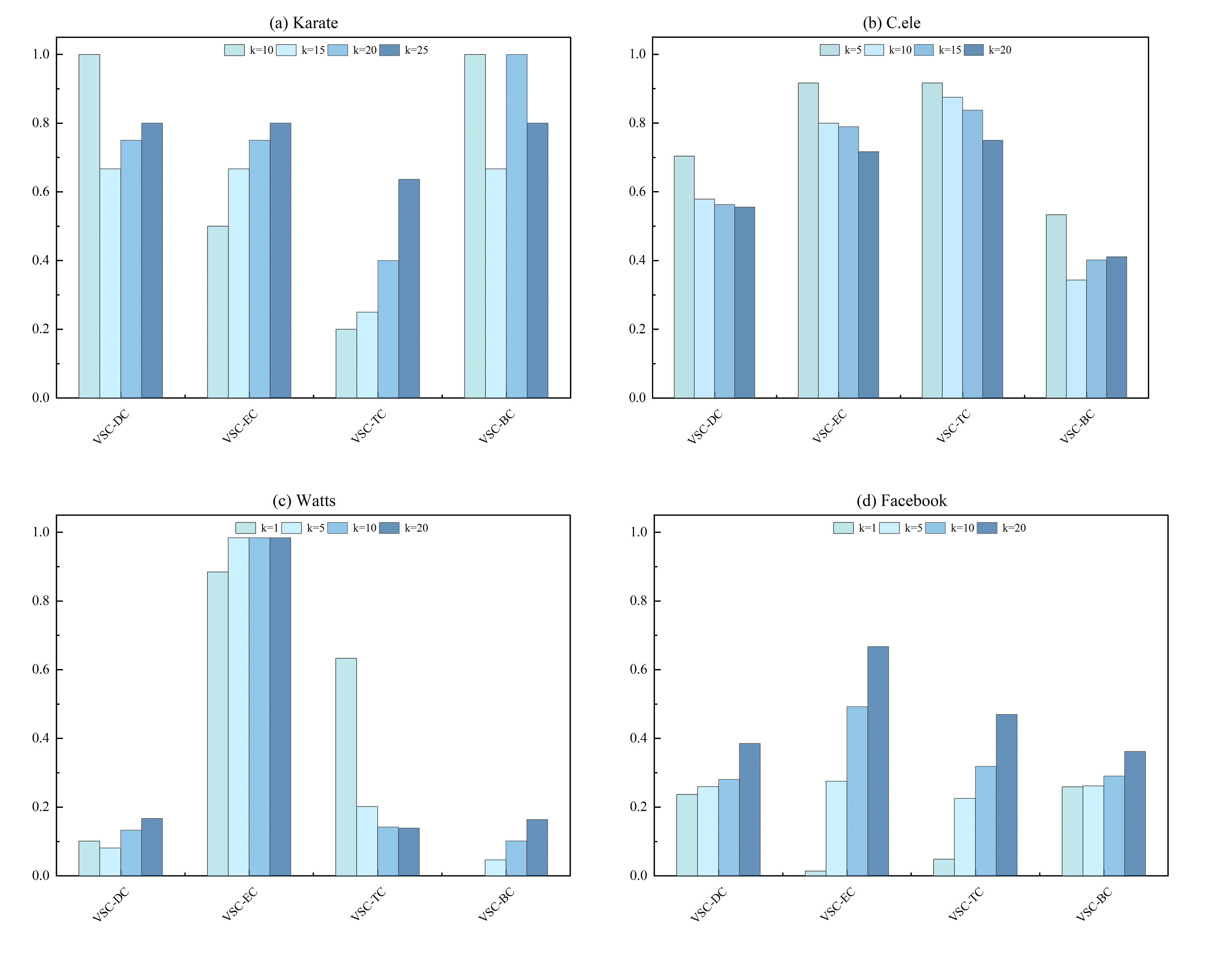}
\caption{Top-$k\%$ Jaccard indices between VSC and the four classical vertex-centrality measures.}\label{fig:jaccard}
\end{figure}

We next examine whether methods select the same \emph{high-ranking} vertices. If $A_k$ and $B_k$ are the sets of Top-$k\%$ vertices selected by two rankings, their Jaccard index is
\[
 J_k(A,B)=\frac{|A_k\cap B_k|}{|A_k\cup B_k|}.
\]
Unlike Spearman's coefficient, which evaluates the relative ordering over all vertices, $J_k$ ignores ranks outside the selected sets and measures only their overlap. The values for VSC and each comparator are shown in Fig.~\ref{fig:jaccard}.

In Karate, the VSC--DC and VSC--BC Jaccard indices are both $1$ at $k=10$, whereas the VSC--EC and VSC--TC indices are $0.50$ and $0.20$. At $k=25$, the overlaps with DC, EC, and BC are all $0.80$, and the overlap with TC is approximately $0.64$. These values contrast with the very high VSC--EC Spearman correlation of $0.980$: agreement over the entire ranking does not guarantee equal agreement among the top vertices.

In C.~elegans, the Top-$5\%$ overlaps of VSC with EC and TC are each approximately $0.91$. At $k=20$, the corresponding indices are approximately $0.72$ and $0.75$. Together with their Spearman coefficients, this indicates comparatively strong agreement for both the global ranking and the leading vertex sets in this network.

In Watts, the VSC--EC Jaccard indices remain close to $1$ across the selection proportions shown, in accordance with the Spearman coefficient reported as $1.000$. In contrast, overlaps of VSC with DC and BC are low; the overlap with TC is more noticeable at $k=1$ but decreases as the selected proportion grows. In Facebook, the VSC--EC Spearman coefficient is $0.950$, yet the Top-$1\%$ Jaccard index is close to zero. Its overlap increases to approximately $0.66$ at $k=20$. This is another instance in which a high global rank correlation coexists with substantial differences among the highest-ranked vertices.

Taken together, Figs.~\ref{fig:spearman} and~\ref{fig:jaccard} provide complementary assessments. VSC is not identical to any of the four existing measures, and the differences depend both on the network and on whether one considers all ranks or only a leading subset. The coupled equations~\eqref{eq:vsc-coupled} explain why such differences may arise: VSC incorporates the scores and capacities of three-vertex subgraphs in addition to vertex-level information.

\subsection{Capacity ablation and ranking changes}\label{subsec:ablation}
To isolate the role of subgraph capacity, we replace $\kappa_\alpha$ by $1$ for every $\alpha\in[m]$, while leaving the nonzero positions of $\mathcal T_G$ unchanged. Denote the resulting tensor by $\widetilde{\mathcal T}_G$. Thus its corresponding nonzero entries are $1/6$ rather than $\kappa_\alpha/6$. As $\mathcal T_G$ and $\widetilde{\mathcal T}_G$ have the same support, $\widetilde{\mathcal T}_G$ is also weakly irreducible. The resulting \emph{capacity-ablated VSC} vector is the unique $4$-norm-normalized positive Perron eigenvector $\widetilde{\mathbf z}=(\widetilde{\mathbf x}^{\mathrm T},\widetilde{\mathbf y}^{\mathrm T})^{\mathrm T}$ of $\widetilde{\mathcal T}_G$. Both sets of scores are recomputed rather than obtained by rescaling the original scores.

Table~\ref{tab:top15_ablation} lists the fifteen highest-ranked vertices and three-vertex subgraphs of each network before and after ablation. The listed subgraph is identified by its three member vertices, and its type is $P_3$ or $K_3$.

\begin{table}[htbp]
\caption{Top-$15$ ranked vertices and induced subgraphs before and after capacity ablation on four networks (Karate and C.~elegans).}\label{tab:top15_ablation}
\centering
\scriptsize
\setlength{\tabcolsep}{3.5pt}
\renewcommand{\arraystretch}{1.13}
\begin{adjustbox}{max width=\textwidth}
\begin{tabular}{c c c c c c c c}
\toprule
\multirow{2}{*}{Network} & \multirow{2}{*}{Rank} & \multicolumn{3}{c}{VSC} & \multicolumn{3}{c}{Capacity-ablated VSC} \\
\cmidrule(lr){3-5} \cmidrule(lr){6-8}
& & Vertex & Subgraph & Type & Vertex & Subgraph & Type \\
\midrule
\multirow{15}{*}{Karate}
& 1 & 34 & $\{9,33,34\}$ & $K_3$ & 34 & $\{3,33,34\}$ & $P_3$ \\
& 2 & 1 & $\{32,33,34\}$ & $K_3$ & 1 & $\{1,9,34\}$ & $P_3$ \\
& 3 & 33 & $\{31,33,34\}$ & $K_3$ & 33 & $\{1,32,34\}$ & $P_3$ \\
& 4 & 3 & $\{1,3,9\}$ & $K_3$ & 3 & $\{9,33,34\}$ & $K_3$ \\
& 5 & 9 & $\{3,9,33\}$ & $K_3$ & 9 & $\{1,14,34\}$ & $P_3$ \\
& 6 & 14 & $\{1,3,14\}$ & $K_3$ & 32 & $\{32,33,34\}$ & $K_3$ \\
& 7 & 32 & $\{24,33,34\}$ & $K_3$ & 14 & $\{3,9,34\}$ & $P_3$ \\
& 8 & 2 & $\{1,2,3\}$ & $K_3$ & 2 & $\{14,33,34\}$ & $P_3$ \\
& 9 & 31 & $\{30,33,34\}$ & $K_3$ & 31 & $\{1,3,33\}$ & $P_3$ \\
& 10 & 20 & $\{9,31,34\}$ & $K_3$ & 20 & $\{3,14,34\}$ & $P_3$ \\
& 11 & 4 & $\{15,33,34\}$ & $K_3$ & 28 & $\{1,20,34\}$ & $P_3$ \\
& 12 & 24 & $\{16,33,34\}$ & $K_3$ & 24 & $\{31,33,34\}$ & $K_3$ \\
& 13 & 28 & $\{19,33,34\}$ & $K_3$ & 29 & $\{20,33,34\}$ & $P_3$ \\
& 14 & 29 & $\{21,33,34\}$ & $K_3$ & 30 & $\{1,9,33\}$ & $P_3$ \\
& 15 & 30 & $\{23,33,34\}$ & $K_3$ & 10 & $\{9,32,34\}$ & $P_3$ \\

\midrule

\multirow{15}{*}{C.~elegans}
& 1  & 186 & $\{147,186,408\}$ & $K_3$ & 186 & $\{147,186,408\}$ & $K_3$ \\
& 2  & 147 & $\{145,147,186\}$ & $K_3$ & 147 & $\{145,147,186\}$ & $K_3$ \\
& 3  & 408 & $\{145,186,408\}$ & $K_3$ & 408 & $\{145,186,408\}$ & $K_3$ \\
& 4  & 145 & $\{147,186,231\}$ & $K_3$ & 145 & $\{147,186,231\}$ & $K_3$ \\
& 5  & 231 & $\{147,186,205\}$ & $K_3$ & 231 & $\{147,186,205\}$ & $K_3$ \\
& 6  & 205 & $\{186,231,408\}$ & $K_3$ & 205 & $\{147,186,227\}$ & $P_3$ \\
& 7  & 228 & $\{186,205,408\}$ & $K_3$ & 228 & $\{147,186,228\}$ & $P_3$ \\
& 8  & 229 & $\{186,228,408\}$ & $K_3$ & 229 & $\{186,231,408\}$ & $K_3$ \\
& 9  & 155 & $\{186,227,408\}$ & $K_3$ & 155 & $\{186,205,408\}$ & $K_3$ \\
& 10 & 146 & $\{147,153,186\}$ & $K_3$ & 146 & $\{186,227,408\}$ & $K_3$ \\
& 11 & 153 & $\{153,186,408\}$ & $K_3$ & 153 & $\{186,228,408\}$ & $K_3$ \\
& 12 & 427 & $\{145,186,231\}$ & $K_3$ & 427 & $\{147,153,186\}$ & $K_3$ \\
& 13 & 295 & $\{145,186,205\}$ & $K_3$ & 295 & $\{153,186,408\}$ & $K_3$ \\
& 14 & 311 & $\{147,186,426\}$ & $K_3$ & 311 & $\{147,186,426\}$ & $K_3$ \\
& 15 & 208 & $\{147,155,186\}$ & $K_3$ & 409 & $\{145,186,231\}$ & $K_3$ \\
\bottomrule
\end{tabular}
\end{adjustbox}
\end{table}

\begin{table}[htbp]
\caption{Top-$15$ ranked vertices and induced subgraphs before and after capacity ablation (Watts and Fb-pages; continued from Table~\ref{tab:top15_ablation}).}\label{tab:top15_ablation_continued}
\centering
\scriptsize
\setlength{\tabcolsep}{3.5pt}
\renewcommand{\arraystretch}{1.13}
\begin{adjustbox}{max width=\textwidth}
\begin{tabular}{c c c c c c c c}
\toprule
\multirow{2}{*}{Network} & \multirow{2}{*}{Rank} & \multicolumn{3}{c}{VSC} & \multicolumn{3}{c}{Capacity-ablated VSC} \\
\cmidrule(lr){3-5} \cmidrule(lr){6-8}
& & Vertex & Subgraph & Type & Vertex & Subgraph & Type \\
\midrule
\multirow{15}{*}{Watts}
& 1 & 4346 & $\{4337,4346,4382\}$ & $K_3$ & 4346 & $\{4337,4346,4382\}$ & $K_3$ \\
& 2 & 4382 & $\{4337,4346,4396\}$ & $K_3$ & 4337 & $\{4337,4346,4396\}$ & $K_3$ \\
& 3 & 4337 & $\{4337,4346,4374\}$ & $K_3$ & 4382 & $\{4346,4382,4396\}$ & $P_3$ \\
& 4 & 4396 & $\{4337,4346,4362\}$ & $K_3$ & 4396 & $\{4337,4346,4374\}$ & $K_3$ \\
& 5 & 4333 & $\{4346,4348,4382\}$ & $K_3$ & 4374 & $\{4346,4374,4382\}$ & $P_3$ \\
& 6 & 4374 & $\{4337,4346,4348\}$ & $K_3$ & 4333 & $\{4333,4337,4346\}$ & $P_3$ \\
& 7 & 4362 & $\{4333,4337,4382\}$ & $K_3$ & 4362 & $\{4333,4346,4382\}$ & $P_3$ \\
& 8 & 4348 & $\{4346,4374,4396\}$ & $K_3$ & 4348 & $\{4337,4346,4362\}$ & $K_3$ \\
& 9 & 4353 & $\{4346,4362,4396\}$ & $K_3$ & 4347 & $\{4346,4374,4396\}$ & $K_3$ \\
& 10 & 4403 & $\{4337,4348,4382\}$ & $K_3$ & 4403 & $\{4346,4362,4382\}$ & $P_3$ \\
& 11 & 4347 & $\{4346,4362,4374\}$ & $K_3$ & 4353 & $\{4337,4382,4396\}$ & $P_3$ \\
& 12 & 4385 & $\{4346,4347,4382\}$ & $K_3$ & 4373 & $\{4346,4362,4396\}$ & $K_3$ \\
& 13 & 4373 & $\{4337,4346,4347\}$ & $K_3$ & 4343 & $\{4337,4346,4348\}$ & $K_3$ \\
& 14 & 4399 & $\{4337,4374,4396\}$ & $K_3$ & 4366 & $\{4337,4374,4382\}$ & $P_3$ \\
& 15 & 4343 & $\{4337,4362,4396\}$ & $K_3$ & 4392 & $\{4346,4348,4382\}$ & $K_3$ \\

\midrule

\multirow{15}{*}{Fb-pages}
& 1 & 8397 & $\{787,1238,8397\}$ & $K_3$ & 8397 & $\{787,8397,11381\}$ & $P_3$ \\
& 2 & 11381 & $\{787,6659,8397\}$ & $K_3$ & 11381 & $\{1238,6659,8397\}$ & $P_3$ \\
& 3 & 11935 & $\{13,1238,8397\}$ & $K_3$ & 11935 & $\{13,8397,11381\}$ & $P_3$ \\
& 4 & 1238 & $\{13,6659,8397\}$ & $K_3$ & 1238 & $\{6476,8397,11381\}$ & $P_3$ \\
& 5 & 8224 & $\{6476,6659,8397\}$ & $K_3$ & 8224 & $\{8397,11381,13636\}$ & $P_3$ \\
& 6 & 6659 & $\{1238,8397,13636\}$ & $K_3$ & 857 & $\{787,1238,8397\}$ & $K_3$ \\
& 7 & 857 & $\{1212,1238,8397\}$ & $K_3$ & 6659 & $\{487,8397,11381\}$ & $P_3$ \\
& 8 & 8925 & $\{1238,2703,8397\}$ & $K_3$ & 1855 & $\{2703,8397,11381\}$ & $P_3$ \\
& 9 & 1855 & $\{6659,8397,13636\}$ & $K_3$ & 8925 & $\{1212,8397,11381\}$ & $P_3$ \\
& 10 & 787 & $\{787,8397,11381\}$ & $P_3$ & 787 & $\{787,8224,8397\}$ & $P_3$ \\
& 11 & 13 & $\{1238,5634,8397\}$ & $K_3$ & 13 & $\{8397,10054,11381\}$ & $P_3$ \\
& 12 & 11096 & $\{6607,6659,8397\}$ & $K_3$ & 3894 & $\{3123,8397,11381\}$ & $P_3$ \\
& 13 & 6476 & $\{3222,6659,8397\}$ & $K_3$ & 6476 & $\{787,6659,8397\}$ & $K_3$ \\
& 14 & 3894 & $\{487,6659,8397\}$ & $K_3$ & 2898 & $\{3329,8397,11935\}$ & $P_3$ \\
& 15 & 2898 & $\{2703,6659,8397\}$ & $K_3$ & 11096 & $\{5634,8397,11381\}$ & $P_3$ \\
\bottomrule
\end{tabular}
\end{adjustbox}
\end{table}

The leading vertex sets are relatively stable under ablation. The original and ablated Top-$15$ lists share $14$, $14$, $13$, and $15$ vertices in Karate, C.~elegans, Watts, and Facebook, respectively. In Karate, the Top-$10$ vertex sets are identical, whereas the order of some entries changes. In C.~elegans, the first fourteen vertices and their order remain unchanged. Some high-ranking vertices exchange positions in Watts and Facebook, showing that substantial overlap need not mean identical ordering or unchanged centrality scores.

The subgraph rankings are more sensitive to the capacity weights. In the original VSC model, the Top-$15$ subgraphs include $15$, $15$, $15$, and $14$ triangles, respectively, in the four networks. After ablation, these counts become $3$, $13$, $8$, and $2$. The largest changes in the type distribution occur in Karate and Facebook, whereas most leading subgraphs of C.~elegans remain triangles even when the capacities are set to one.

This difference follows from the subgraph component of~\eqref{eq:vsc-coupled}:
\begin{equation}\label{eq:ablation-subgraph}
 \rho(\mathcal T_G)y_\alpha^3
 =\kappa_\alpha\prod_{i\in V_\alpha}x_i.
\end{equation}
The original model assigns factors $2$ and $3$ to induced paths and triangles, whereas the ablated model assigns the same factor to both. In addition, the member-vertex scores are affected by this change through the coupled vertex equations. Capacity therefore affects subgraph scores both directly and through the jointly determined vertex scores. The observed ablation results are consistent with a stronger influence on the composition of the leading subgraph lists than on the membership of the leading vertex lists.

\subsection{Triangle proportions among high-ranking subgraphs}\label{subsec:triangles}
\begin{figure}[h]
\centering
\includegraphics[width=0.8\linewidth]{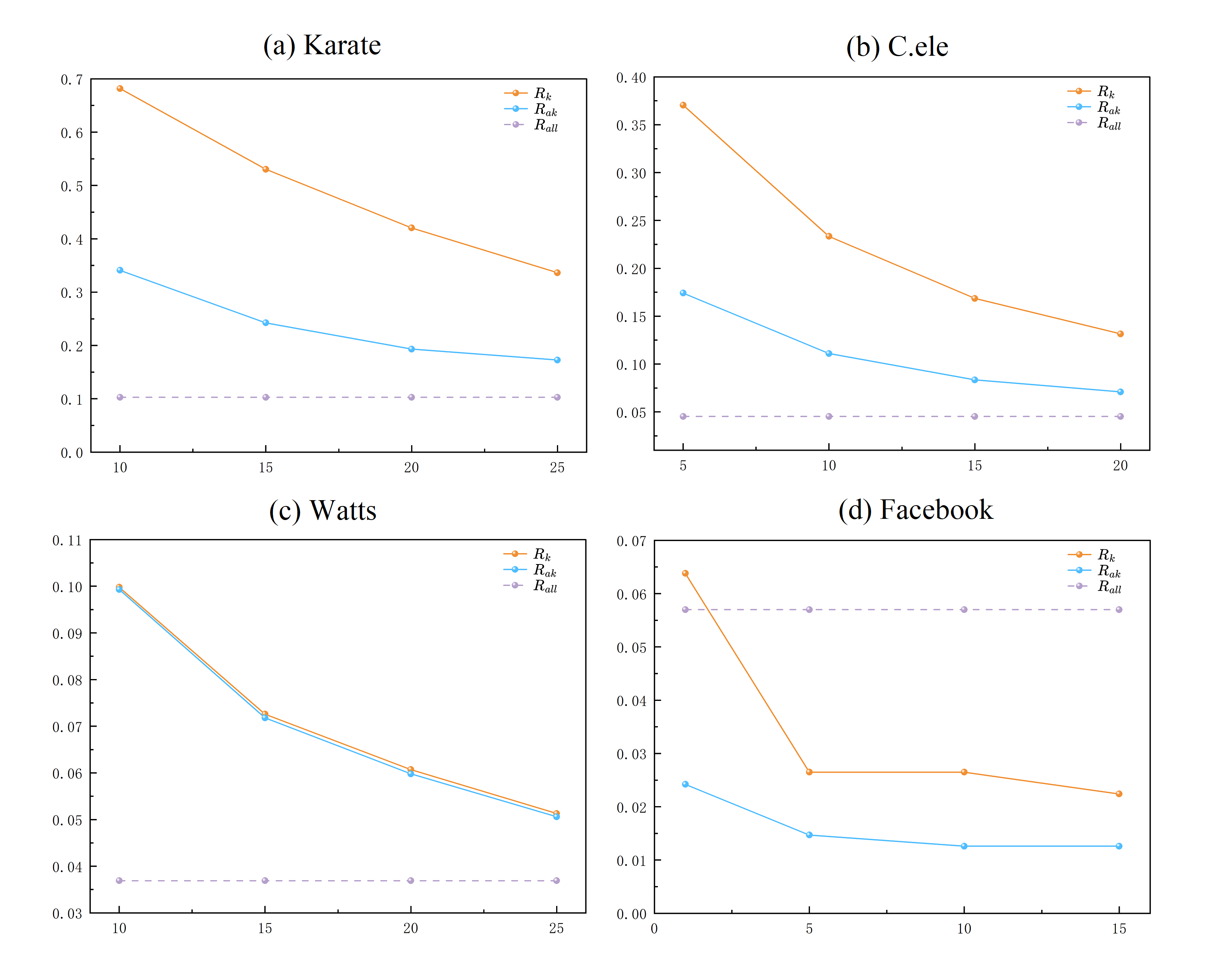}
\caption{Triangle proportions in the Top-$k\%$ three-vertex subgraphs, with and without capacity weighting. The overall triangle proportion $R_{\mathrm{all}}$ is included as a baseline.}\label{fig:triangle-ratio}
\end{figure}

To examine larger selections, let $\mathcal S_k,\widetilde{\mathcal S}_k\subseteq[m]$ be the index sets of the Top-$k\%$ three-vertex subgraphs under VSC and capacity-ablated VSC, respectively. Define
\begin{equation}\label{eq:triangle-ratios}
\begin{aligned}
 R_k&=\frac{|\{\alpha\in\mathcal S_k:G[V_\alpha]\cong K_3\}|}{|\mathcal S_k|},\\[3pt]
 R_{ak}&=\frac{|\{\alpha\in\widetilde{\mathcal S}_k:G[V_\alpha]\cong K_3\}|}{|\widetilde{\mathcal S}_k|}, \\[3pt]
  R_{\mathrm{all}}&=\frac{m_{K_3}}{m}.
\end{aligned}
\end{equation}
Here $R_k$ and $R_{ak}$ are the triangle proportions in the selected sets and $R_{\mathrm{all}}$ is the corresponding proportion among all three-vertex subgraphs. Figure~\ref{fig:triangle-ratio} compares these quantities across selection proportions.

For the selection proportions shown, the triangle proportion under VSC is generally higher than under capacity ablation, although the gap depends on the network and on $k$. In Karate and C.~elegans, the difference is especially apparent at small selection proportions. In Watts, the curves are relatively close over much of the displayed range, despite the change among the first fifteen subgraphs. In Facebook, capacity weighting also produces a larger triangle proportion for small $k$, while the triangle proportions of both models decrease as the selected proportion increases. These trends should be interpreted relative to the different baseline proportions $R_{\mathrm{all}}$ in Table~\ref{tab:network}.

The Top-$15$ rankings and the Top-$k\%$ comparisons address complementary questions. The former identify concrete subgraphs whose positions change; the latter quantify triangle representation in larger selected sets. Together, they indicate that capacity weighting affects the relative ranking of $P_3$ and $K_3$ subgraphs. The extent of this effect also depends on the network's distribution of highly ranked member vertices. These results do not imply that triangles must outrank induced paths in every graph.

\section{Conclusion}\label{sec:conclusion}
We proposed vertex--subgraph eigenvector centrality (VSC) to jointly evaluate the centralities of vertices and connected induced subgraphs of order three. The coupled equations combine the scores of the three member vertices, the score of the induced subgraph, and its capacity, defined as the number of internal edges. These equations admit an exact fourth-order symmetric nonnegative tensor representation. We proved that this tensor is weakly irreducible for every connected graph with at least three vertices, ensuring a uniquely normalized positive eigenvector.

A four-vertex example makes the mutual reinforcement and tensor entries explicit. Numerical experiments show that the vertex rankings generated by VSC have network-dependent relationships with degree, eigenvector, triangle, and betweenness centralities. In particular, high Spearman correlation does not necessarily imply high overlap among the highest-ranked vertices. Capacity-ablation experiments further indicate that the weighting has a more visible effect on leading induced-subgraph rankings, including their proportions of triangles, than on the membership of the leading vertex sets in the networks considered. An extension to connected induced subgraphs on more than three vertices is a natural direction for future work.
\section*{CRediT authorship contribution statement}
\textbf{Qingying Zhang:} Methodology, writing--original draft, visualization, validation.
\textbf{Haotian Zeng:} Investigation, writing--original draft.
\textbf{Changjiang Bu:} Conceptualization, writing--review and editing.
All authors have read and approved the final version of the manuscript.

\section*{Declaration of competing interest}
The authors declare that they have no known competing financial interests or personal relationships that could have appeared to influence the work reported in this paper.

\section*{Acknowledgments}
The research of the second author is partially supported by the National Natural Science Foundation of China (No.~12071097) and the Natural Science Foundation for the Excellent Youth Scholars of Heilongjiang Province (No.~YQ2022A002).

\section*{References}
\bibliographystyle{unsrt}
\bibliography{ml0ht2}

\end{spacing}
\end{document}